# Multiscale Computation of a Polypeptide Backbone Model

**Dov Bai**

Department of Chemistry
University of Cincinnati
Cincinnati, OH 45221-0172
Email: dbai@juno.com

## 1. Abstract

The multiscale Monte-Carlo algorithm outlined in Bai and Brandt[1] is applied to a simple model of the polypeptide backbone. Effective coarse level Hamiltonians are derived by a fast Newtonian iterative scheme. The coarse Hamiltonian parameters are adjusted so that local structural properties have the same value in both coarse and fine level simulations. It is demonstrated that at convergence of iterations, global structural properties are reproduced very well in coarse level simulations.

## 2. Introduction

In this work, the multiscale Monte-Carlo (MC) algorithm presented in Bai and Brandt[1] for polymers is applied to a simple model of the polypeptide backbone. More details on the application to polymeric systems are a subject of another article. The purpose of studying the simple model is to provide some starting point for applying the algorithm to complex proteins. The simple model includes only backbone interaction terms and no side chains. Two cases are studied: The first includes bonding terms and dipole-dipole interactions. The second adds Van der Waals forces by including WCA potential terms. While hydrogen bond terms are included too, the initial configuration is in a coil state and no transitions to helix were found.

The general philosophy in Bai and Brandt[1] and similarly in the context of Algebraic Multigrid (AMG) methods[16,18] and Renormalization Multigrid (RMG)[13] is this: It is assumed that the parameters of the coarse Hamiltonian depend on local properties of the fine chain and thus can be derived from short chains. The parameters are derived by requiring good agreement of local structural properties in fine and coarse simulations. The local structural properties may include, for example, probability densities of coarse internal coordinates and correlations between neighboring internal coordinates. Similar to Bai and Brandt[1], it is demonstrated that once the locally dependent Hamiltonian parameters are derived, an excellent fit of end-to-end probability densities is achieved. Others, such as Fukunaga et al.[4] also identified the importance of local correlations, but little or no attempt was made to include them explicitly in coarse-grained Hamiltonians.

In recent years considerable progress was made in the development of coarse-graining procedures for long molecular chains. For a comprehensive review see Müller-Plathe[17]. Some[3,4] provide examples of deriving coarse bonding terms and expressed them as functions of coarse internal coordinates while nonbonded interactions are derived from radial distribution functions (RDF). Generally, however, the authors do not provide a systematic method for relating measurable observables (averages over configurations of some quantity) to Hamiltonian terms or criteria for selecting the terms to include in the coarse Hamiltonian. Notably, most of the terms are derived separately, and although useful as first order approximation of the coarse Hamiltonian, no general way is provided for deriving higher order terms, which reflect interactions between first order terms.

The structure of this article is as follows: In Section 3, the details of the model are provided. Section 4 discusses the coarsening procedure. Section 5 provides information on fine level simulations. It demonstrates the typical behavior of the modeled chain in the presence of specific interaction terms. By studying the coarse properties of fine chains such as probability densities of coarse dihedral angles, it is possible to get a better qualitative understanding of the chain structure.

Section 6 contains most of the results. The first part reviews the iterative process of deriving coarse level Hamiltonians. It also describes the two main types of observables that are selected for adjusting coarse Hamiltonian parameters. Sections 6.1 and 6.2 provide details of the application to the backbone chain. In Section 6.1 the interactions include bonding and nearest-neighbor dipole-dipole terms and in Section 6.2 WCA potential terms[9] are added. Finally, Section 7 includes some concluding remarks regarding the choice of the model and the suitability of the model and the algorithm for studying more complex phenomena of such as protein folding.

## 3. The model

The simple model consists of a sequence of residues. The $i$-th residue includes the three consecutive backbone atoms $N_i$, $C_i^a$, $C_i$ (Hollow circles in Figure 1). No side chains are included. When a hydrogen bond potential term between residues $i$ and $j$ is added, the locations of $H(N_i)$ and $O(C_j)$, where $H(N_i)$ is the hydrogen atom bonded to $N_i$ and $O(C_j)$ is the oxygen atom bonded to $C_j$ are calculated. However, $H(N_i)$ and $O(C_j)$ do not participate in any other interaction.

The Hamiltonian includes the following interactions:

I.  Bonding forces:
    A. Bond-length term between any two adjacent backbone atoms:
    $$E_l = K_{a_1,a_2}(r_{a_1,a_2} - L_{a_1,a_2})^2 \qquad (1)$$
    where $a_1, a_2$ can be any of the following pairs of successive atoms: $(C_i^a, C_i)$, $(C_i, N_{i+1})$ or $(N_i, C_i^a)$, $r_{a_1,a_2}$ is the distance between the two atoms of the pair and $K_{a_1,a_2}$, $L_{a_1,a_2}$ are constants having the following numerical values:
    $L_{C^a,C}$=1.53Å, $L_{C,N}$ =1.32Å, $L_{N,C^a}$ =1.45Å, $K_{C^a,C}$=317 Kcal/mol Å$^{-2}$,
    $K_{C,N}$ =490 Kcal/mol Å$^{-2}$, $K_{N,C^a}$ =337 Kcal/mol Å$^{-2}$.
    B. Bond-angle term between any three consecutive backbone atoms:
    $$E_a = K_{a_1,a_2,a_3}(a_{a_1,a_2,a_3} - A_{a_1,a_2,a_3})^2 \qquad (2)$$
    where $(a_1, a_2, a_3)$ can be any of the triplets of successive atoms:
    $(C_i^a, C_i, N_{i+1})$, $(C_i, N_{i+1}, C_{i+1}^a)$ or $(N_i, C_i^a, C_i)$, $a_{a_1,a_2,a_3}$ is the angle formed by

the atoms and $K_{a_1,a_2,a_3}$, $A_{a_1,a_2,a_3}$ are constants having the following numerical values: $A_{C^a,C,N} = 115°$, $A_{C,N,C^a} = 121°$, $A_{N,C^a,C} = 109°$, $K_{C^a,C,N} = 70$ Kcal/mol degrees$^{-2}$, $K_{C,N,C^a} = 50$ Kcal/mol degrees$^{-2}$, $K_{N,C^a,C} = 80$ Kcal/mol degrees$^{-2}$.

C. Torsion term between any four succesive backbone atoms: $(C_i, N_{i+1}, C^a_{i+1}, C_{i+1})$ ($\varphi$), $(N_i, C^a_i, C_i, N_{i+1})$ ($\psi$) and $(C^a_i, C_i, N_{i+1}, C^a_{i+1})$ ($\omega$). A torsion angle can have values in the interval $[-180°, 180)$ (or $[-\pi, \pi)$ radians) where the *trans* configuration is at $180°$ ($\pi$ radians). For $\varphi$ and $\psi$ the potential term is:

$$E_t = 1/2 K_t (1 + \cos 3t) \qquad (3)$$

where $t = \varphi$ or $\psi$ and $K_\varphi = K_\psi = 1$ Kcal/mol (Chap. VII in Flory[5]). The torsion angle $\omega$ is nearly frozen at the *trans* state - A potential term $K_\omega (1 + \cos \omega)$ is added for $\omega$ where the constant $K_\omega$ is set arbitrarily to the large value of 1000 Kcal/mol and enables fluctuations of about $5°$ around the *trans* state.

II. Dipole-dipole interaction:

Following the discussion in Brant and Flory[7] and others[5,6,8], a dipole moment is assigned at the midpoint of each amide bond, forming an angle of $56°$ with it. The magnitude of the dipole moment is 3.7D (or 0.72 $e$Å, where $e$ is the electron charge). The interaction term between two point dipoles is:

$$E_d = \frac{332}{\epsilon} \left[ \mu_A \cdot \mu_B / r^3 - 3(\mu_A \cdot r)(\mu_B \cdot r) / r^5 \right], \qquad (4)$$

where $E_d$ is expressed in units of Kcal/mol, $\mu_A$, $\mu_B$ are the dipole moments expressed in units of $e$Å, $r$ is a vector from point A to point B and $\epsilon$ is the relative dielectric constant. Only nearest-neighbor dipole-dipole interactions are included and $\epsilon = 3.5$ [7,8].

III. Van der Waals (vdw) forces:

It is assumed that due to polypeptide-solvent interactions there is no attractive part of vdw forces[7]. Thus, the following WCA potential[9] is used:

$$E^{i_B j_B}_{WCA}(r_{i_B j_B}) = F_{i_B j_B} \begin{cases} 0 & r_{i_B j_B} > 2^{1/6} \sigma \\ V_0 \left[ (\sigma / r_{i_B j_B})^{12} - (\sigma / r_{i_B j_B})^6 + \frac{1}{4} \right] & r_{i_B j_B} \leq 2^{1/6} \sigma \end{cases} \qquad (5)$$

where $i_B$ denotes the index of the atom (rather then its residue) along the backbone, $r_{i_B, j_B}$ is the distance between two atoms with indices $i_B$ and $j_B$ and $F_{i_B j_B}$ is given by:

$$F_{i_B j_B} = \begin{cases} 0 & |i_B - j_B| < 3 \\ 1/2 & |i_B - j_B| = 3 \\ 1 & |i_B - j_B| > 3. \end{cases} \qquad (6)$$

For C,C pairs: $s = 3.21$Å, $e = 0.556$ Kcal/mol, for C,N pairs $s = 3.07$Å, $e = 0.683$ Kcal/mol and for N,N pairs $s = 2.94$Å, $V_0 = 0.854$ Kcal/mol.

IV. Hydrogen bonds:

In order to detect transitions between coil and $a$-helix configurations a hydrogen bond term is added between any donor residual $i$ and acceptor residual $i - 4$. The hydrogen bond term is[10-12]:

$$E_{hb} = \frac{A_{hb}}{r_{N-O}^{12}} - \frac{B_{hb}}{r_{N-O}^{10}}, \qquad (7)$$

where $r_{N-O}$ is the distance between the donor's nitrogen and the acceptor's oxygen atoms. The constants $A_{hb}$ and $B_{hb}$ in Eq. (7) are selected so that $E_{hb}$ has a minimum of $-1.11$ Kcal/mol at $r_{N-O} = 2.9$Å.

## 4. Coarsening

In Bai and Brandt[1], the coarsening of a simple polymethylene chain is done by taking geometric averages of groups of consecutive atoms. Coarsening here is done in a similar way: The fine level is divided into groups of 3 backbone atoms and the geometric center of each group is a coarse point. Figure 1 describes a piece of a backbone chain with 3 residues $i, i+1, i+2$ and their atoms (hollow circles). The 3 black circles, $P_i, P_{i+1}, P_{i+2}$ are coarse level points. $P_i$ is the geometric average of atoms $N_i$, $C_i^a$ and $C_i$. $P_{i+1}$ and $P_{i+2}$ are calculated in a similar way.

The effectiveness of coarsening schemes is tested by *compatible Monte-Carlo* (CMC) iterations. For details on CMC see, for example, Brandt and Ron[13] or Brandt and Ilyin[14]. The purpose of CMC iterations is to verify that it is possible to produce efficiently the fine level configurations corresponding to a given coarse state. The decorrelation times of "slow" variables (variables with higher decorrelation times than others) are tested. If their decorrelation times are small, it is an indication of a good coarsening scheme. Note that CMC decorrelation times are independent of chain size.

Bai and Brandt[1] discuss CMC iterations. As will be published elsewhere, numerical experiments revealed that the coarsening ratio of 1:4 (one coarse point per 4 fine points) used there for polymethylene, is inadequate, as it did not prohibit transitions between the two *gauche* and *trans* states of the fine level dihedral angles. This is corrected by changing the coarsening ratio to 1:3. Similarly, in this study, the coarsening ratio is set to 1:3 as no transitions between dihedral states were detected in long running CMC tests. While here a coarsening ratio of 1:4 was found to be acceptable too due to nearly freezing of the $w$ torsion angles, a coarsening ratio of 1:3 is more natural, as there are 3 backbone atoms per residue.

## 5. Fine level simulations

Fine level Monte-Carlo simulations are the starting point for deriving the coarse Hamiltonian. These simulations were done on a backbone chain consisting of 21 residues (63 backbone atoms). The initial chain is in a random coil state and does not contain $a$-helices.

All simulations are at $k_BT = 0.5$ Kcal/mol. The Metropolis algorithm is used. The moves are such that one internal coordinate at a time is changed randomly and the new configuration is accepted or rejected. A single MC iteration consists of 3 sweeps over the entire chain: In the first sweep each bond length is changed, one at a time. In the second, each bond angle is changed and finally in the third pass each dihedral angle is changed. The decorrelation times for central dihedral angles $\varphi$ and $\psi$ were between 1 and 20, depending on the interaction terms included. To create a database of statistically independent configurations for use in deriving the coarse Hamiltonian, configurations created after $Di$, $i = 0,1,2,...$ iterations, where $D$ is the decorrelation time, were saved in the database. Typically the database consists of 200000 fine configurations.

Three cases are considered:

1) Chains with bonding forces only.
2) Chains with bonding forces and nearest neighbor dipole-dipole interactions.
3) Chains with bonding forces, nearest neighbor dipole-dipole interactions, Van der Waals forces and hydrogen bonds. As mentioned earlier no $\alpha$-helices were detected in the simulations and therefore the hydrogen bond terms have little effect.

Section 3 provides details of the interactions.

In Figure 2 a typical chain produced by simulations with bonding forces only (case 1) is plotted and in Figure 3 a typical chain with added dipole-dipole interaction terms (case 2). The chain in Figure 2 is similar to a typical polymethylene chain with bonding forces only which is discussed in Section 2 of Bai and Brandt[1] and detailed results of its two-level coarsening are provided there. Although the coarsening ratio in Bai and Brandt[1] is 1:4, because of the general similarity no numerical results with coarse level Hamiltonian are presented here.

Figures 5 and 6 describe the probability densities of central coarse bond-length and bond-angle for case 2 as measured in fine level simulations. Figure 4 describes the probability densities of the central coarse dihedral for cases 1 and 2. With the absence of dipole-dipole interactions, the coarse dihedral has nearly a uniform distribution. With dipole-dipole interactions there is a strong preference for the coarse dihedrals to be at or near the *trans* configuration. Indeed, by comparing Figures 2 and 3, it is apparent that adding dipole-dipole interactions significantly increases the average size of the chain as discussed in Brant and Flory[6]. As all internal coarse dihedrals have similar distributions, and longer chains have greater tendency for *trans* configurations, Figure 4 provides information on the global structure of the chains.

Figures 7-9 describe the probability densities of central coarse coordinates as measured in fine simulations for a chain with WCA and hydrogen bond forces added (case 3). Comparing Figs. 4-6 and 7-9 it is clear that the length and angle distributions without WCA potential are sharper and that dihedral configurations with WCA are less clustered around the *trans* state than in case 2, indicating higher probabilities for shorter chains.

## 6. Coarse Hamiltonians

The first part of this section reviews the process of derivation of the coarse Hamiltonian. In Section 6.1 a coarse Hamiltonian is derived for a backbone with bonding and nearest neighbor dipole-dipole interactions (case 2 of Section 5), and in Section 6.2 Van der Waals and hydrogen bond potential terms are added (case 3 of Section 5).

The coarse Hamiltonian $H^c$ is expressed as a sum of $N_H$ Hamiltonian terms:

$$H^c = \sum_{k=1}^{N_H} A_k H_k. \qquad (8)$$

The $H_k$ are functions of coarse coordinates and the $A_k$ are Hamiltonian parameters. The parameters are adjusted in an iterative process so that local structural properties have the same values in fine and coarse simulations.

In the following discussion the notation $\langle \cdot \rangle$ is used to denote an average over all configurations. $\langle H_k \rangle$ is the average over all configurations of the Hamiltonian term $H_k$ and $\langle H_k H_l \rangle$ is the average of the multiplications of two such terms, $H_k$ and $H_l$. Denote by $\langle H_k \rangle_f$ and $\langle H_k H_l \rangle_f$ the first and second order observables measured in simulations with the fine level Hamiltonian, and by $\langle H_k \rangle_c$ and $\langle H_k H_l \rangle_c$ the same observables measured with the coarse Hamiltonian. In Bai and Brandt[1] it is shown that by using a procedure similar to Swendsen[2] for the Ising model and later by Lyubartsev and Laaksonen[19-20] for different potentials, the corrections to the coarse Hamiltonian parameters is obtained by solving the linear system:

$$\langle H_k \rangle_f - \langle H_k \rangle_c = \frac{1}{k_B T} \sum_{l=1}^{N_H} \big(\langle H_k \rangle \langle H_l \rangle - \langle H_k H_l \rangle \big) dA_l, \qquad k = 1, \ldots N_H \qquad (9)$$

where $k_B$ is Boltzmann's constant, $T$ is the temperature, and $dA_l$ is the correction to the Hamiltonian parameter $A_l$. Thus, at the $i$-th iteration, coarse level simulations are conducted with a set of $N_H$ Hamiltonian parameters in $H^c$ and the first and second order observables are measured. Then a new set Hamiltonian parameters is calculated by solving Eq. (9) and updating the Hamiltonian parameters. The updated parameters are used in the coarse Hamiltonian of the $i+1$-th iteration.

In the numerical results that follow, the coarse Hamiltonian includes at least two types of terms: (1) Terms resulting from the probability densities of coarse internal coordinates (bond lengths, bond angles and dihedral angles) and (2) terms which include correlations between internal coordinates. The first type is discussed in detail at the beginning of Section 6.1. The second type is included so that observables like $C_{LL}$, which expresses the average of correlations between adjacent coarse lengths and is given by:

$$C_{LL} = \left\langle \sum_i \big(L_i - \langle L_i \rangle\big)\big(L_{i+1} - \langle L_{i+1} \rangle\big) \right\rangle \qquad (10)$$

where $L_i$ is the $i$-th coarse bond length formed by the coarse points $i$ and $i+1$ (e.g. the distance between $P_i$ and $P_{i+1}$ in Figure 1) and the summation is over all adjacent pairs of lengths, have the same values in both fine and coarse simulations at the convergence of the iterative process.

In addition to $C_{LL}$, the following 5 correlation observables are included:

$$C_{AL} = \left\langle \sum_i \sum_{j=i,i+1} (A_i - \langle A_i \rangle)(L_j - \langle L_j \rangle) \right\rangle \qquad (11)$$

$$C_{AA} = \left\langle \sum_i (A_i - \langle A_i \rangle)(A_{i+1} - \langle A_{i+1} \rangle) \right\rangle \qquad (12)$$

where $A_i$ is the angle formed by the 3 consecutive coarse points $i$, $i+1$, $i+2$ (e.g. $P_i, P_{i+1}, P_{i+2}$ of Figure 1), $L_i$ and $L_{i+1}$ are the two coarse lengths forming $A_i$ and the summation is over all pairs of adjacent angles. The next 3 correlation terms involve the coarse dihedrals $D_i$ (the notation $\overline{D_i}$ is discussed in the next paragraph) formed by 4 consecutive coarse points $i$, $i+1$, $i+2$, $i+3$:

$$C_{DL} = \left\langle \sum_i (\overline{D_i} - \langle \overline{D_i} \rangle)(L_{i+1} - \langle L_{i+1} \rangle) \right\rangle \qquad (13)$$

is the average of the sum of correlations between a dihedral angle and its center length,

$$C_{DA} = \left\langle \sum_{i, j=i,i+1} (\overline{D_i} - \langle \overline{D_i} \rangle)(A_j - \langle A_j \rangle) \right\rangle \qquad (14)$$

is the average of the sum of correlations between a dihedral angle and the two angles forming it, and finally

$$C_{DD} = \left\langle \sum_i (\overline{D_i} - \langle \overline{D_i} \rangle)(\overline{D_{i+1}} - \langle \overline{D_{i+1}} \rangle) \right\rangle \qquad (15)$$

is the average of the sum of correlations between two adjacent dihedral angles.

One has to be careful about correlations involving dihedral angles. Expressed in radians, the range of values of the dihedral is $[-\pi, \pi)$, where $\pi$ is the *trans* angle. However, the dihedral angle is discontinuous at the *trans* angle and the term $D_i - \langle \overline{D_i} \rangle$ changes sign there. To correct, $D_i$ is transformed to $\overline{D_i}$ before calculating correlations involving the dihedral angles.

For the chain with bonding and nearest neighbor dipole-dipole interaction terms, the probability density of the dihedral angles (Figure 4) has a sharp peak near *trans* but vanishing probabilities near *cis*. Therefore, the transformation from $D_i$ to $\overline{D_i}$ is defined by:

$$\overline{D_i} = \begin{cases} D_i + 2\pi & -\pi \leq D_i < 0 \\ D_i & \pi > D_i \geq 0. \end{cases} \qquad (16)$$

When the WCA potential is included the probability of dihedral angles near *cis* is different from 0 (Figure 9) and the transformation (16) can no longer be used because of the discontinuity of $\overline{D}_i - \langle \overline{D}_i \rangle$ near the *cis*. Instead, the transformation

$$\overline{D}_i = \begin{cases} (D_i + 2\mathbf{p})(-D_i/\mathbf{p}) & -\mathbf{p} \leq D_i < 0 \\ D_i(D_i/\mathbf{p}) & \mathbf{p} > D_i \geq 0 \end{cases} \quad (17)$$

is used. The transformation (17) is similar to (16) but is multiplied by a "tent" function centered at $\langle \overline{D}_i \rangle$ and is zero at $\overline{D}_i = 0, 2\mathbf{p}$.

## *6.1. Bonding and dipole-dipole forces*

The local observables that are selected to have the same values in simulations with coarse and fine Hamiltonians are:

1) Probability density of all internal coordinates: The interval of values for each internal coordinate was divided into 16 subintervals. The size of each subinterval is set such that the probability of the internal coordinate to be found in any of the subintervals is approximately the same. The reason for the approximate equi-probability division is to avoid near zero terms in the linear system (9).
2) The six correlation terms of Eqs. (10)-(16). These were discussed above.

The probability densities of central internal coordinates are plotted in Figures 4-6. Table I lists the values of the 6 correlation terms. Clearly the dihedral-dihedral ($C_{DD}$) is the dominant correlation.

The $H_k$ in Eq. (8) consist of terms corresponding to the observables. As suggested in Bai and Brandt[1], in order to increase the accuracy of the probability density and thus to enable a smaller number of subintervals, a linear rather then constant interpolation is used to approximate probability densities: If the total interval of an internal coordinate $U_i$ is $I_i = [X_0^i, X_{n_i}^i]$ where $X_0^i < X_1^i < ... X_{n_i}^i$ is a grid placed over $I_i$ dividing it into $n_i$ subintervals then the Hamiltonian term for the $j$-th subinterval is:

$$H_{i,j} = \begin{cases} 0 & \text{if } U_i \leq X_{j-1}^i \\ \dfrac{U_i - X_{j-1}^i}{X_j^i - X_{j-1}^i} & \text{if } X_{j-1}^i \leq U_i \leq X_j^i \\ \dfrac{X_{j+1}^i - U_i}{X_{j+1}^i - X_j^i} & \text{if } X_j^i \leq U_i \leq X_{j+1}^i \\ 0 & \text{if } X_{j+1}^i \leq U_i. \end{cases} \quad (j = 1,...,n_i - 1) \quad (18)$$

For the 21 coarse points the total number of internal coordinates is 57 (20 lengths + 19 angles + 18 dihedral angles). As each coordinate is divided into 16 subintervals and for each subinterval there is a Hamiltonian term $H_{i,j}$ (Eq. 18), the total number of terms of the first type is 912=16 · 57. For each correlation in Eqs. 10-15 there is a corresponding

Hamiltonian term – the term inside the angle brackets. For example, the Hamiltonian term corresponding to $C_{LL}$ (Eq. 10) is:

$$H_{LL} = \sum_i (L_i - \langle L_i \rangle)(L_{i+1} - \langle L_{i+1} \rangle) \tag{19}$$

and similarly for $H_{AL}$, $H_{AA}$, $H_{DL}$, $H_{DA}$ and $H_{DD}$. Thus, the total number of Hamiltonian terms is 918=912+6. In the following discussion the *coordinate parameter* $A_{i,j}$ is the Hamiltonian parameter multiplying the Hamiltonian term $H_{i,j}$ (Eq. 18). The *correlation parameters* $A_{LL}$, $A_{LA}$, $A_{AA}$, $A_{DL}$, $A_{DA}$ and $A_{DD}$ are the Hamiltonian parameters multiplying the Hamiltonian terms $H_{LL}$, $H_{LA}$, $H_{AA}$, $H_{DL}$, $H_{DA}$ and $H_{DD}$. As a first approximation to the coarse Hamiltonian all correlation parameters are set to 0, and the coordinate parameters are given by:

$$A_{i,j} = -k_B T \ln \langle H_{i,j} \rangle_f / |I_{i,j}| \tag{20}$$

Where $\langle H_{i,j} \rangle_f$ is the fine average of $H_{i,j}$ and $|I_{i,j}|$ is the size of the subinterval corresponding to $H_{i,j}$. With that approximation, simulations with $H^c$ yield $\langle H_{i,j} \rangle_f = \langle H_{i,j} \rangle_c$ and zero correlations, i.e. reproduce the probability density of internal coordinates but all coarse correlations are vanishing.

To check the effect of coarse Hamiltonian parameters on global structural quantities, the end-to-end (ETE) probability density was measured in coarse simulations and compared to the fine. The solid line in Figure 10 describes the ETE probability density as measured on the fine level. The dotted line is the probability density when doing coarse level simulations with the first approximation Hamiltonian. Typically, about 180000 MC coarse iterations similar to those described in Section 5 were made to measure the coarse observables.

Since $C_{DD}$ is much larger than the other 5 correlations, the first numerical experiments where done with $A_{LL} = A_{LA} = A_{AA} = A_{DL} = A_{DA} = 0$ and changing only $A_{DD}$. The coordinate parameters $A_{i,j}$ were kept the same as in the first Hamiltonian approximation (i.e. Eq. (20)). Thus only a single Hamiltonian parameter, $A_{DD}$, was updated in the iterations. The iterations stopped when the observed coarse dihedral-dihedral correlation calculated by the right-hand-side of Eq. (15) (denoted here by $C_{DD}^c$ - The superscript $c$ is used to indicate correlations that are measured with the coarse Hamiltonian ) was about the same as the fine value of –2.61 (Table I). The dependence of $C_{DD}^c$ on $A_{DD}$ is highly non-linear. At convergence $A_{DD}$=0.42 and $C_{DD}^c$=-2.70. The dashed line in Figure 10 describes the ETE probability density at convergence. Clearly, just by adjusting a single Hamiltonian parameter such that the dominant local correlation has the same values on coarse and fine resulted in highly improved global structural properties.

Next, numerical experiments were done in which all 6 coarse correlation parameters, rather then just $A_{DD}$ were updated in each iteration. The iterative scheme described at the beginning of this section was used. The iterations stopped when all 6 correlations measured in coarse simulations were about the same as those obtained on the fine level and are listed in Table I. Table II lists the correlations obtained at the convergence of the iterative process. Comparing Table II and Table I, all 6 coarse correlations are nearly identical to the fine values. The resulting ETE probability density is similar to the dotted line in Figure 10. Details of the convergence of the iterative process are omitted here as a more general case is described in detail in the next section.

In the above numerical experiments the 912 averages $\langle H_{i,j}\rangle_c$ were measured in the iterations and no significant deviations from $\langle H_{i,j}\rangle_f$ were found. Thus, it was not necessary to correct the coordinate parameters in the iterative process.

## *6.2. Bonding, dipole-dipole and WCA forces*

In addition to the two types of terms of Section 6.1, coarse WCA potential terms need to be added to the coarse Hamiltonian. In principle, a third type:

3) Averages of $E_{WCA}^{IJ}(R_{IJ})$

should be included where $E_{WCA}^{IJ}(R_{IJ})$ is a function of the distance $R_{IJ}$ between the two coarse points $I$ and $J$. On the fine level $E_{WCA}^{IJ}(R_{IJ})$ is calculated by

$$E_{WCA}^{IJ}(R_{IJ}) = \sum_{i_B \in I, j_B \in J} E_{WCA}^{i_B j_B}(r_{i_B j_B}) \qquad (21)$$

where the summation is over all fine points $i_B$ and $j_B$ "belonging" to the coarse points $I$ and $J$ (e.g. $N_i$, $C_i^a$ and $C_i$ in Figure 1 "belong" to the coarse point $P_i$) and $E_{WCA}^{i_B j_B}(r_{i_B j_B})$ is given by Eq. (5). For coarsening ratio of 1:3, the summation in Eq. (21) includes 9 terms. $R_{IJ}$ may be "binned" and a pair of coarse Hamiltonian term and coefficient may correspond to each bin. The added Hamiltonian coefficients due to the WCA potential should be included in the linear system (9) as part of the iterative process.

However, it was found in numerical experiments that the fine averages of $E_{WCA}^{IJ}(R_{IJ})$ can be approximated by an analytic function, which has a similar form to the fine WCA potential (Eq. (5)). When this function is included in the coarse Hamiltonian and multiplied by a constant coefficient equals 1, at convergence the coordinate and correlation parameters are rather insensitive to small perturbations in the shape of this function. It is therefore appropriate to assume that it is not necessary to parameterize the coarse WCA potential. In the numerical results that follow, the coarse WCA is given by (5) with $r_{i_B j_B}$ replaced by $R_{IJ}$, $s$ =4.187Å, $V_0$=0.1 Kcal/mol and $F_{i_B j_B}$ is replaced by $F_{IJ}^c$ defined by:

$$F_{IJ}^c = \begin{cases} 0 & |I-J|<2 \\ 1 & |I-J|\geq 2. \end{cases} \qquad (22)$$

Table III lists the 6 fine level correlations. Similar to the correlations of the previous section listed in Table I, the dihedral-dihedral correlation is the dominant one. However, the angle-length ($C_{AL}$) and angle-angle ($C_{AA}$) correlations are considerably larger than their counterparts in Table I.

The coordinate and correlation parameters in the first approximation of the Hamiltonian are the same as in Section 6.1: All correlation parameters are set to 0 and the coordinate parameters $A_{i,j}$ are given by Eq. (20). Note that coarse level simulations with the first approximation Hamiltonian do not necessarily result in vanishing correlations, because of the WCA potential.

Figure 11 describes the ETE probability density. The solid line is the probability density of the fine level. The peak of the distribution occurs at a smaller distance than in Figure 10. This is consistent with the differences between coarse dihedral distributions with and without WCA potentials as noted at the end of Sec. 5.

The dashed line is the probability density obtained in simulations with the first approximation coarse Hamiltonian. The dotted line is the probability density at the convergence of the iterative process.

The first step in constructing the coarse Hamiltonian includes MC iterations with the first approximation to the Hamiltonian. Then, several steps follow in which $A_{DD}$ is gradually increased and all other Hamiltonian coefficients remain unchanged. The increase in $A_{DD}$ stops when the observed coarse correlation, $C_{DD}^c$, is close to the fine value. Then, several Newtonian iteration steps are made with all 918 coefficients corrected by solving Eq. (9). The convergence of the iterations is described in Tables IV-V.

The first column in Table IV lists the Newtonian iteration number. The second column lists $C_i$, the factor that multiplies the vector of corrections $dA_l$ obtained by solving Eq. (9). In the first few iterations this value is smaller than 1 in order to avoid divergence because of the non-linearity. Columns 3-5 describe the $L_\infty$ norm (largest absolute value) of the error in each of the 3 types of internal coordinates (length, bond angles, dihedral angles). The last column lists the $L_\infty$ error of the 6 observed correlations.

Table V lists the six observed correlations as a function of the iteration number. The last row lists the fine values, which are the same as in Table III.

It is clear that there is a fast reduction in the error, and no more than 5 Newtonian iterations are needed to reduce the error to the level of the roundoff error.

## 7. Conclusions

In this paper a first attempt was made to extend the multiscale algorithm of Bai and Brandt[1], which was originally developed for polymers, to protein systems. The choice of model for this study was deliberate: on the one hand the backbone of polypeptides with no side chains has resemblance to polymers, which makes it easier both methodically and programmatically to apply the algorithm. On the other hand, the model includes many of the force field terms of a real protein and thus also many of their computational characteristics. It is demonstrated that the same multiscale algorithmic principle of using locally derived coarse Hamiltonian parameters leads to accurate global structural properties on the coarse level.

The initial configuration of the Monte-Carlo simulations in this study is in the coil state. In spite of the inclusion of hydrogen-bond terms, no transition to helix configuration was detected in fine level simulations. However, separate fine level numerical experiments with the same model, initially in $a$ - helix configuration, show that the helix configuration is stable, and there were no transitions from the helix to coil states. This leads to the conclusion that in the future, by designing an appropriate coarsening scheme and cycling between fine and coarse level configurations, it may be possible to use the same or similar models for developing fast protein folding algorithms. It should be noted that coarsening schemes which are verified by compatible Monte-Carlo (CMC) tests have "build-in" efficient transitions from the coarse to fine levels (Section 4).

## 8. Acknowledgements

The research was supported by a Department of Defense/Army Innovative Research Grant. The author thanks Achi Brandt, Thomas Beck and Ruth Pachter for the fruitful discussions and their encouragement and support of this research.

## 9. References

1. Bai, D.; Brandt, A. NATO Science Series III, 177, IOS Press, 2001, pp. 250-266.
2. Swendsen, R. H. Phys Rev Lett 1979, 42, 859-861.
3. Reith D.; Pütz M.; Müller-Plathe F. J Comput Chem 2003, 24, 1624-1636.
4. Fukunaga, H; Takimoto, J.; Doi M. J Chem Phys 2002, 116, 8183-8190.
5. P. J. Flory, Statistical Mechanics of Chain Molecules; Interscience; 1969.
6. Brant D. A.; Flory P. J. J Am Chem Soc 1965, 87, 663-664.
7. Brant D. A., Flory P. J. J Am Chem Soc 1965, 87, 2791-2800.
8. Cantor and Schimmel, Biophysical Chemistry: Part I. The conformation of biological macromolecules; W. H. Freeman; 1980.
9. Allen M. P.; Tildesley D. J., Computer simulation of liquids; Oxford University Press; Oxford, 1987.
10. McGuire, R. F.; Momany, F. A.; Scheraga ,H. A. J Phys Chem 1972, 76, 375-393.
11. Momany, F. A.; McGuire, R. F.; Burgess, A. W.; Scheraga H. A. J Phys Chem 1975, **79**, 2361-2380.
12. Momany, F. A; Carruthers L. M.; McGuire R. F.; Scheraga, H. A. J Phys Chem 1974, **78**, 1595-1620.
13. Brandt, A.; Ron D. J Stat Phys 2001, 102, 231-257.
14. Brandt, A.; Ilyin, V. Journal of Molecular Liquids 2003, 105, 245-248.
15. Tschop, W.; Kremer, K.; Batoulis, J.; T. Berger, T.; Hahn, O. Acta Polymer 1998, 49, 61-74.
16. J. W. Ruge, K. Stuben, *Algebraic Multigrid*, in Multigrid Methods, SIAM Frontiers in Applied Mathematics, 1987.
17. Müller-Plathe F. Chem Phys Chem 2002, 3, 754-769.
18. Brandt, A. Electronic Trans Num Anal 2000, 10, 1-10.
19. Lyubartsev, A. P.; Laaksonen, A. Phys Rev E 1995, 52, 3730-3737.
20. Lyubartsev, A. P.; Laaksonen, A. J Phys Chem 1999, 111, 11207-11215.

| $C_{LL}$ | $C_{AL}$ | $C_{AA}$ | $C_{DL}$ | $C_{DA}$ | $C_{DD}$ |
|---|---|---|---|---|---|
| 9.994(-3) | -4.043(-2) | -9.418(-2) | -2.709(-3) | -4.122(-3) | -2.608 |

**Table I Correlations for bonding and dipole-dipole interactions**

| $C^c_{LL}$ | $C^c_{AL}$ | $C^c_{AA}$ | $C^c_{DL}$ | $C^c_{DA}$ | $C^c_{DD}$ |
|---|---|---|---|---|---|
| 1.101(-2) | -4.057(-2) | -9.360(-2) | -4.377(-3) | -2.871(-2) | -2.580 |

**Table II Correlations obtained on coarse, bonding and dipole-dipole interactions**

| $C_{LL}$ | $C_{AL}$ | $C_{AA}$ | $C_{DL}$ | $C_{DA}$ | $C_{DD}$ |
|---|---|---|---|---|---|
| 0.04 | -0.64 | -0.17 | -0.01 | -0.01 | -1.19 |

**Table III Correlations for bonding, dipole-dipole and WCA potentials**

| Iteration | $C_i$ | $E_l$ | $E_a$ | $E_d$ | $E_{corr}$ |
|---|---|---|---|---|---|
| 1 | 0.5 | 0.1126 | 0.10000 | 0.005091 | 1.318 |
| 2 | 0.5 | 0.0682 | 0.05954 | 0.003769 | 0.8354 |
| 3 | 0.5 | 0.0455 | 0.03884 | 0.003776 | 0.5172 |
| 4 | 1.0 | 0.0244 | 0.02724 | 0.002615 | 0.3054 |
| 5 | 1.0 | 0.01234 | 0.01838 | 0.004411 | 0.1668 |

**Table IV Errors in observable for the iterative Hamiltonian derivation**

| Iteration | $C^c_{LL}$ | $C^c_{AL}$ | $C^c_{AA}$ | $C^c_{DL}$ | $C^c_{DA}$ | $C^c_{DD}$ |
|---|---|---|---|---|---|---|
| 1 | 0.22 | -1.44 | 0.66 | -0.01 | 0.04 | -1.31 |
| 2 | 0.16 | -1.14 | 0.35 | 0.01 | 0.02 | -1.25 |
| 3 | 0.11 | -0.94 | 0.13 | 0.01 | 0.01 | -1.24 |
| 4 | 0.08 | -0.81 | -0.05 | 0.02 | 0.01 | -1.19 |
| 5 | 0.05 | -0.70 | -0.13 | 0.04 | -0.01 | -1.20 |
| Fine | 0.04 | -0.64 | -0.17 | -0.01 | -0.01 | -1.19 |

**Table V Observed correlations for the iterative Hamiltonian derivation**

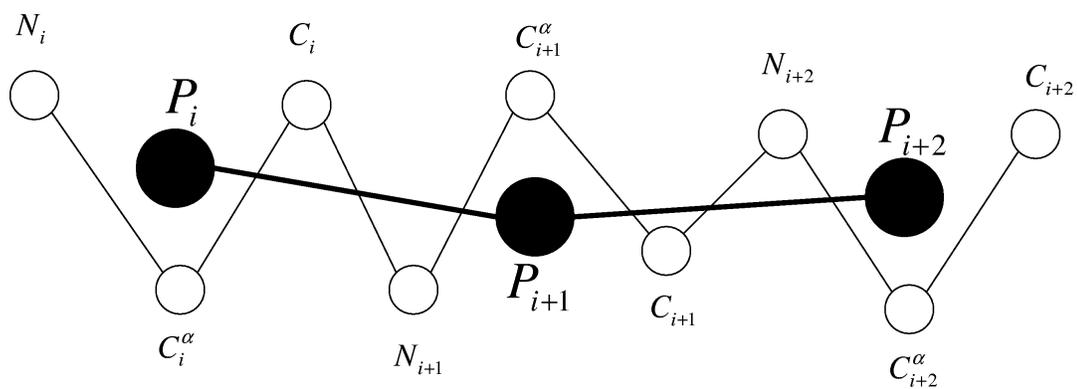

**Figure 1 Fine and coarse levels**

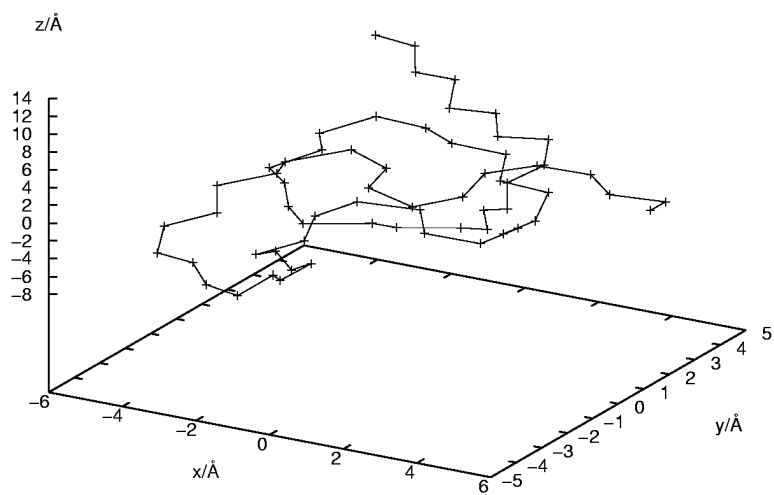

**Figure 2 Chain with bonding forces only**

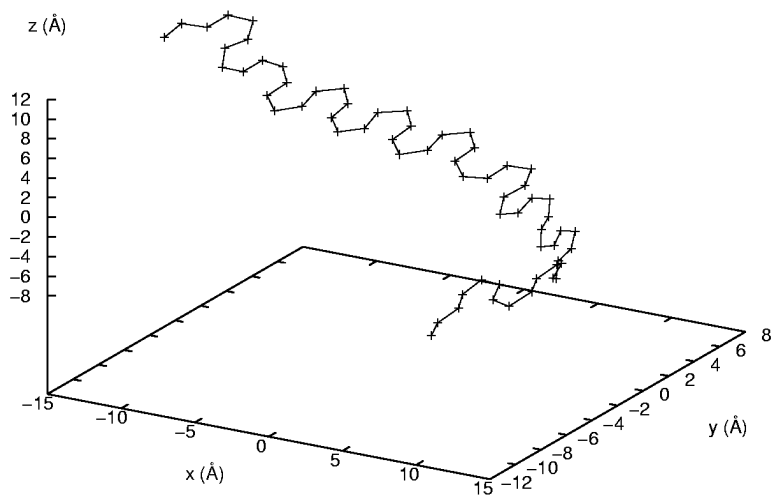

Figure 3 Chain with bonding and dipole-dipole

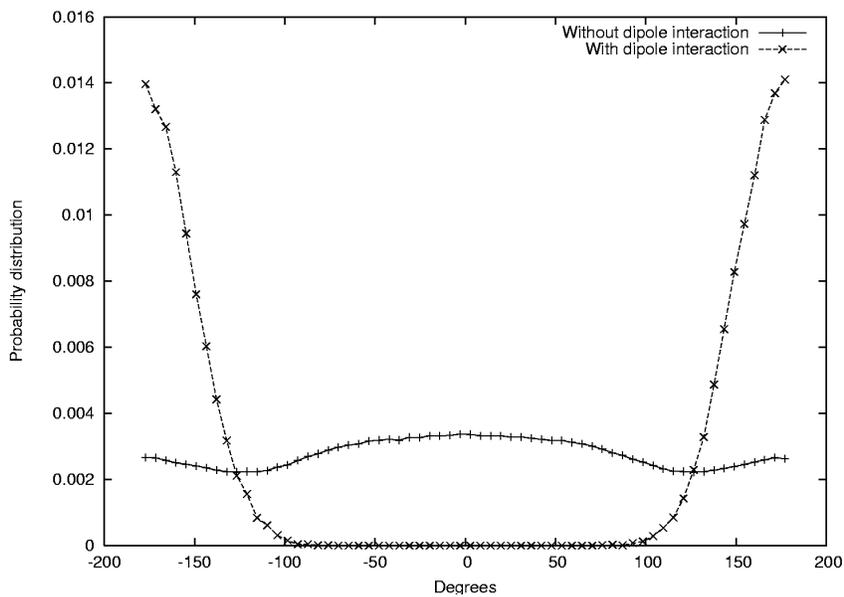

Figure 4 Probability density of coarse dihedrals

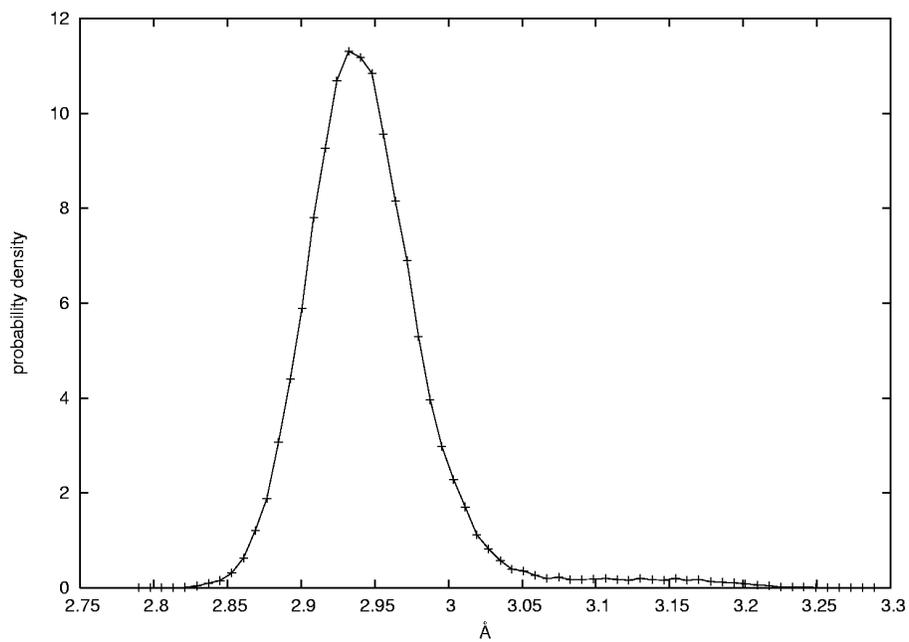

Figure 5 central length (dipole)

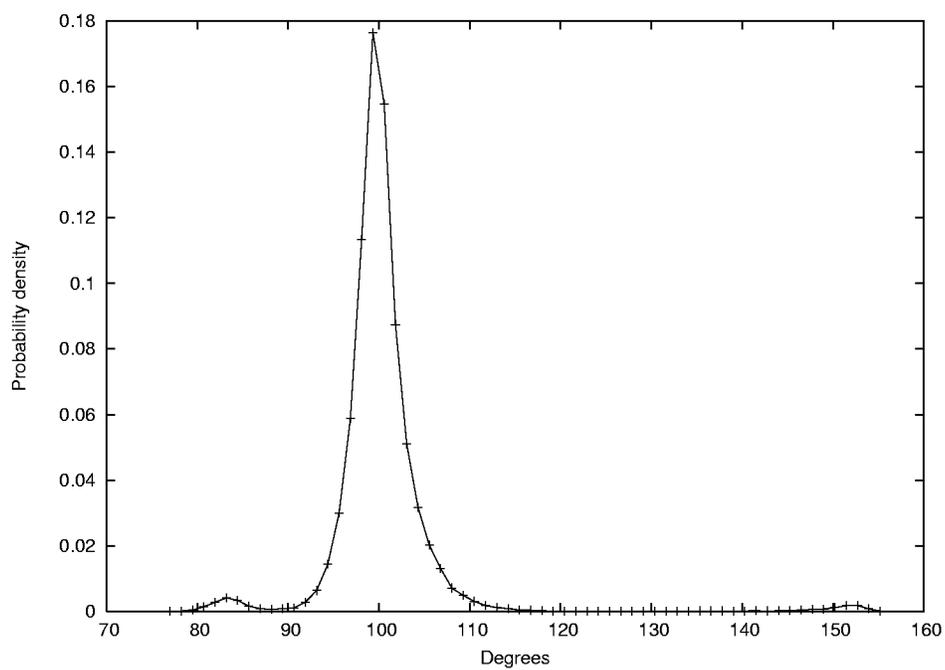

Figure 6 Central angle (dipole)

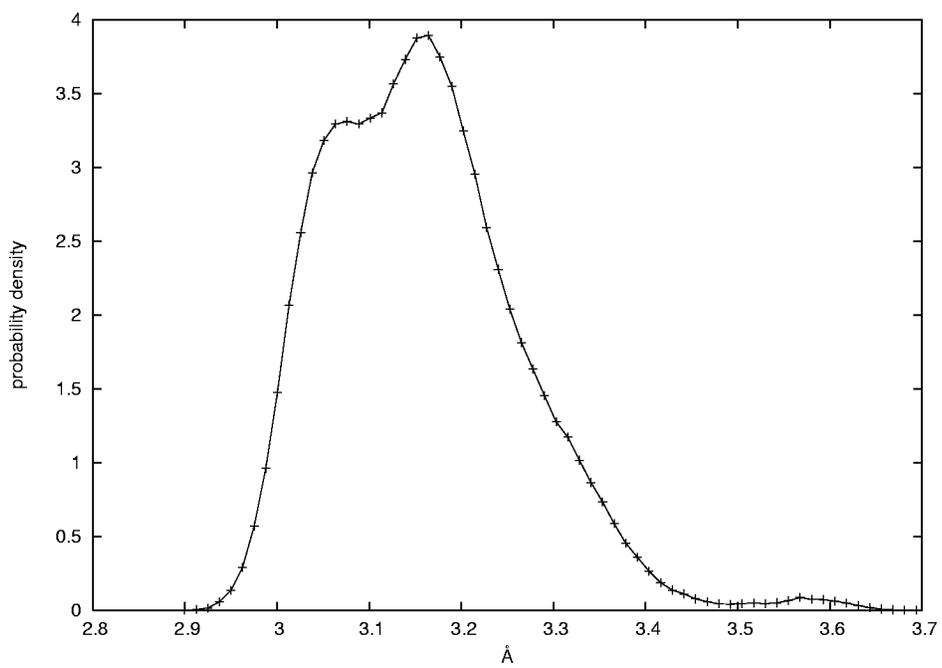

Figure 7 Central length (WCA)

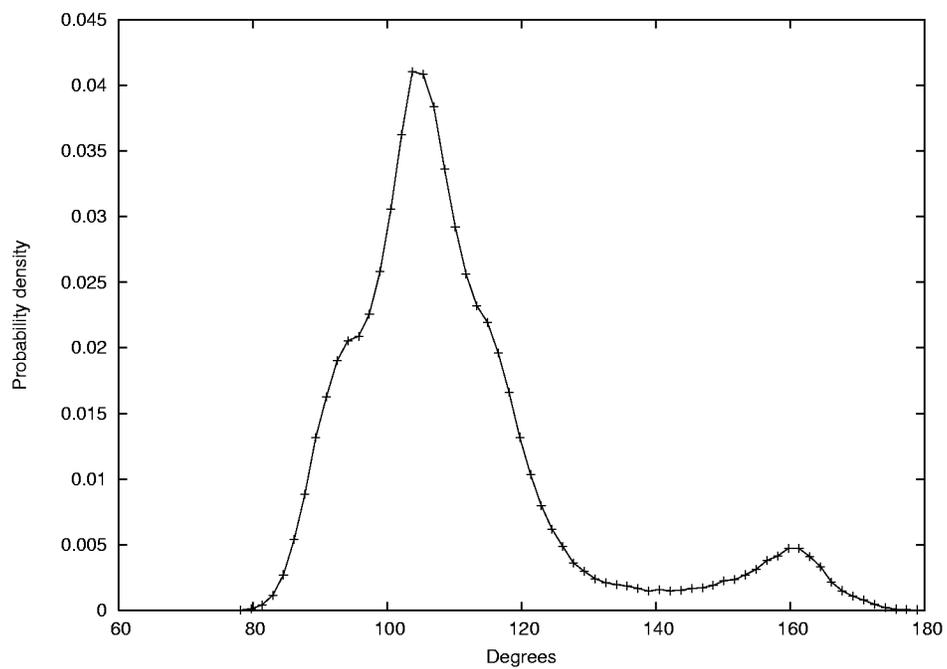

Figure 8 Central angle (WCA)

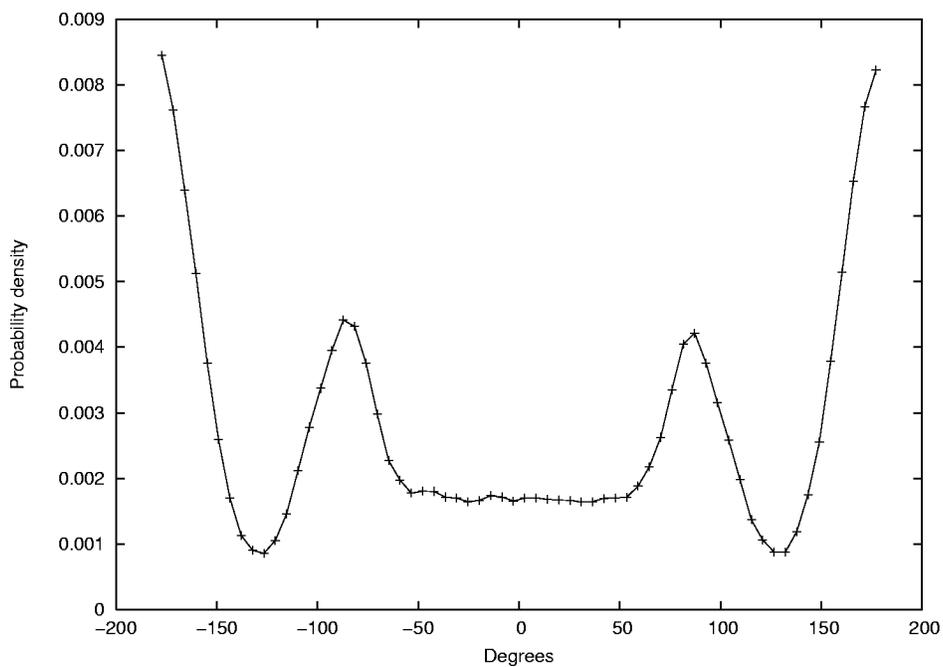

Figure 9 Central dihedral (WCA)

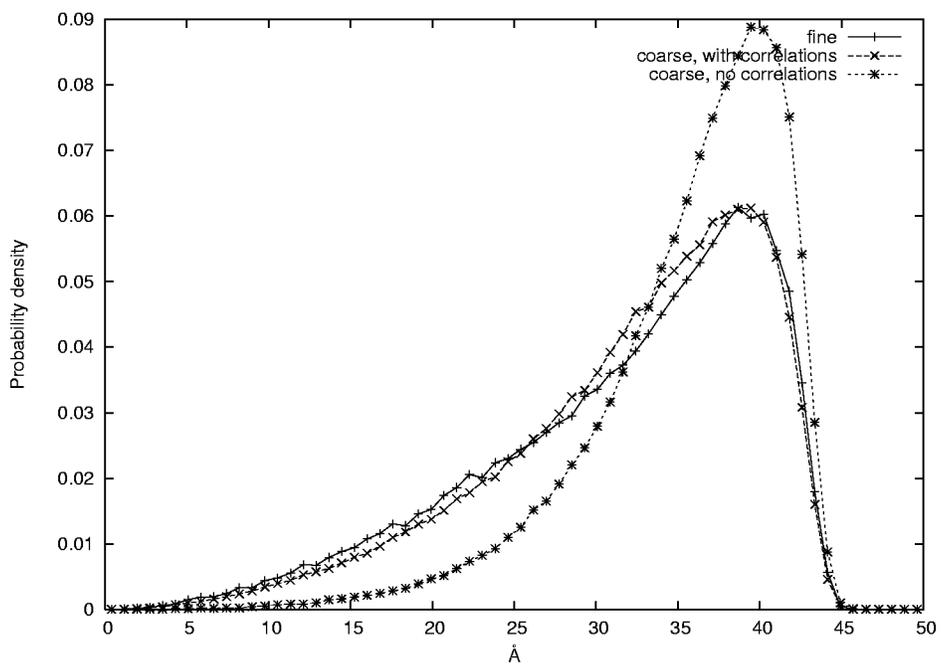

Figure 10 End-to-end distributions, bonding and dipole-dipole interactions

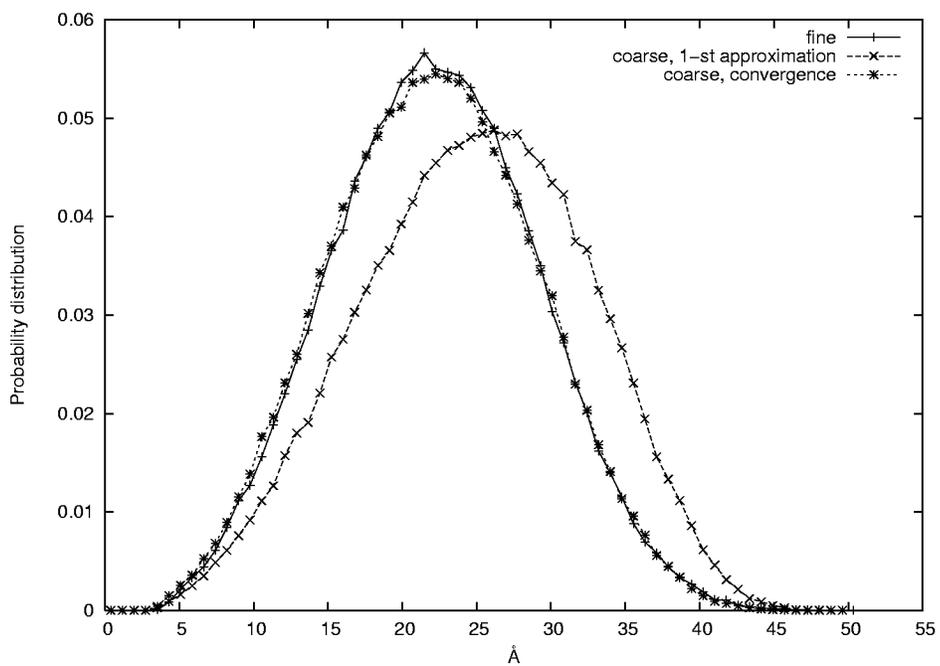

Figure 11 End-to-end distributions, bonding, dipole-dipole and WCA potentials